\documentclass[sigconf,authorversion]{acmart}

\usepackage{booktabs} %

\copyrightyear{2022}
\acmYear{2022}
\setcopyright{rightsretained}
\acmConference[SAC '22]{The 37th ACM/SIGAPP Symposium on Applied Computing}{April 25--29, 2022}{Virtual Event, }
\acmBooktitle{The 37th ACM/SIGAPP Symposium on Applied Computing (SAC '22), April 25--29, 2022, Virtual Event, }\acmDOI{10.1145/3477314.3507165}
\acmISBN{978-1-4503-8713-2/22/04}

\begin{document}
\title{A Priority-Aware Multiqueue NIC Design \\ for Real-Time IoT Devices}
  
\renewcommand{\shorttitle}{A Priority-Aware Multiqueue NIC Design}

\author{Ilja Behnke, Philipp Wiesner, Robert Danicki, Lauritz Thamsen}
\affiliation{%
  \institution{Technische Universität Berlin}
}
\email{{i.behnke,wiesner,r.danicki,lauritz.thamsen}@tu-berlin.de}

\renewcommand{\shortauthors}{Behnke, Wiesner, Danicki, Thamsen}

\begin{abstract}
Low-level embedded systems are used to control cyber-phyiscal systems in industrial and autonomous applications. They need to meet hard real-time requirements as unanticipated controller delays on moving machines can have devastating effects. Modern developments such as the industrial Internet of Things and autonomous machines require these devices to connect to large IP networks. 
Since Network Interface Controllers (NICs) trigger interrupts for incoming packets, real-time embedded systems are subject to unpredictable preemptions when connected to such networks.

In this work, we propose a priority-aware NIC design to moderate network-generated interrupts by mapping IP flows to processes and based on that, consolidates their packets into different queues.
These queues apply priority-dependent interrupt moderation.%
First experimental evaluations show that 93\,\% of interrupts can be saved leading to an 80\,\% decrease of processing delay of critical tasks in the configurations investigated.

\end{abstract}


\maketitle

\section{Introduction}
\label{sec:introduction}
In the context of cyber-physical systems, where software processes lead to physical actions, industrial processing systems and machines are controlled by microcontrollers. %
These devices usually provide only little processing power and run Real-Time Operating Systems (RTOSs) that implement guarantees regarding reaction times for sensory input and control commands~\cite{finn2018introduction}.
With the advent of the Internet of Things (IoT) and Industry 4.0, many cyber-physical systems are being connected to IP networks for remote control, monitoring, and maintenance~\cite{burg2017wireless,lasi2014industry}. %
While the overall architecture of these connected devices is similar to general-purpose microcontrollers, they also have to provide Network Interface Controllers (NICs) and network stack implementations. 

To notify the CPU of newly available data, input devices like sensors and serial interfaces use interrupt requests that trigger Interrupt Service Routines (ISRs) on the controller. 
Due to the distinct priority spaces of ISRs and scheduled processes in RTOSs, interrupts preempt any currently running process independent of its priority~\cite{mejia-alvarez2018interrupt}. 
On real-time devices this introduces a high level of unpredictability as the processing units are interrupted from outside the RTOS~\cite{irq_mngmnt_rt}.
As a result, interrupts triggered by a NIC for incoming network packets and the load-dependent overhead of driver and networking tasks can alter the assumed time-line of running processes. This can result in breaking real-time guarantees for critical processes~\cite{behnke2020interrupting, danicki2021detecting}.%
Consequentially, common IoT devices are easily overwhelmed by network packet floods, even when no subsequent processing of packets is performed. 

Due to this problem, programmers have to resort either to turning off interrupts during critical executions or using separate resources for networking and processing in the typically resource-constrained environments of microcontrollers~\cite{mandalari2021blocking, niedermaier2019secure}.
However, a different solution has to be found in scenarios where IP networks are used for the control of cyber-physical systems and are necessary for critical machine functions.

In this paper, we present the design and preliminary evaluation of a priority-aware multiqueue NIC design that can reduce network-induced interrupt overloads of IoT devices without delaying packets for time-critical tasks. 
This is achieved by reducing the number of interrupts generated by packets related to low priority tasks. The acceptable rate of interrupts can be configured through the operating system to a per-process window.

\section{Multiqueue NIC Design}
\label{sec:approach}
To address the real-time violating effects of network-generated interrupts and their processing overhead, we propose a priority-aware network interface controller that handles network packets depending on their destination process: A configurable multi-queue NIC for real-time embedded systems. This section illustrates the NIC's design, configuration parameters, and operating system-side management.

\begin{figure}
\centering
\includegraphics[width=\linewidth]{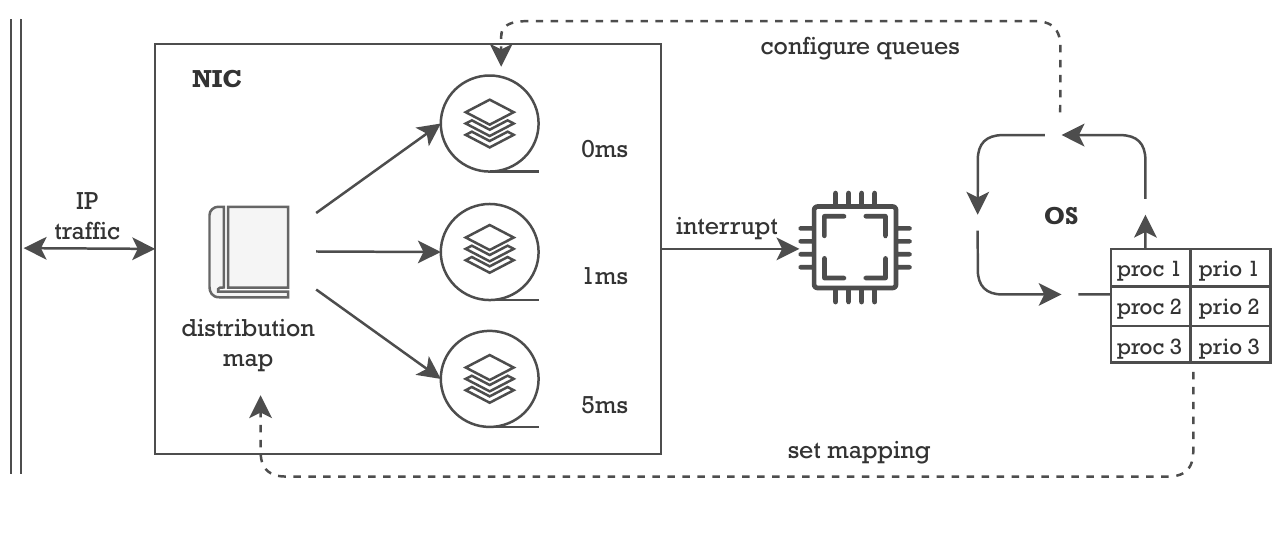}
\caption{Multiqueue NIC: Traffic is organized into different queues with exemplary delay values attached.}%
\label{fig:design}
\end{figure}

\subsection{Heterogeneous Interrupt Moderation}
The issue of network generated interrupts impacting system performance can be addressed using interrupt moderation. However, traditional techniques have their disadvantages. While they increase the overall interrupt processing efficiency, they also increase the incurred packet delays and make them less predictable as packets are held back for a variable amount of time. %
Hence, our NIC design  attempts at reducing the network overhead while guaranteeing low and constant latency for critical packets, i.e. packets used to control critical tasks. 

As embedded systems run a fixed set of specific tasks that is seldomly altered, we can use their metadata to filter and manage incoming packets on the hardware layer before interrupting. Namely, these are the priorities of the packet-receiving processes and their associated IP flows. Interrupt moderation can thereby become the tool to enforce process priorities before entering the operating system domain.

\subsection{NIC Adaptation}
\label{sec:adaptation}
Our proposed NIC adaptation begins after the MAC-layer tasks. An illustration of the design can be seen in Figure \ref{fig:design}. To accommodate incoming packets belonging to different real-time processes, the receive buffer is divided into multiple queues realized as ring buffers. This way, packet descriptors are assigned to different queues depending on their destination process and its priority. 

When a data frame arrives from the network, it is validated and the packet metadata compared to a list of registered ports residing in a distribution map on the NIC. Here, packets are assigned to queues which hold packets of one process each. According to the process priority and expected packet load, different interrupt moderation configurations (e.g. delay timers and counter threshold) are applied to them via the operating system. 

This way, packets for critical processes trigger interrupts immediately upon reception while less important packets (that is, packets with low priority receiving tasks) are held back before one interrupt is triggered for all packets in the respective queue, indicated by the millisecond specifications in Figure \ref{fig:design}. Packets with no associated process can be dropped before an interrupt is triggered, preventing unnecessary processing. 
This is especially important with high unanticipated traffic loads targeting the device and potentially leading to a denial of service. %

\subsubsection*{Relevant Parameters}
The multiqueue NIC introduces four main parameters affecting packet delays and resource utilization:%

\begin{itemize}
\item \emph{Number of queues.} The number of queues the receive buffer is divided into is the number of processes accepting packets.
\item \emph{Size of queues.} The size of a queue corresponds to its expected packet load, available memory, and moderation parameters.
\item \emph{Absolute queue timer values.} Queue-specific periodic duration until an interrupt is triggered by the queue.
\item \emph{Packet timer values.} Queue-specific interrupt timer that is being reset reset by each incoming packet.
\end{itemize}

The timer values are used to span a time window of how long a packet remains in the queue. Depending on the packet rate, a variable number of packets is then coalesced for one interrupt. 
As these parameters have a high impact on the timeliness of incoming traffic and generated workload on the real-time device, the accuracy of their configuration is of high importance.

\subsubsection*{Configuration}
As depicted in Figure \ref{fig:design}, there are two interfaces necessary for NIC configurations. One for the tuning of the before mentioned queue parameters and secondly, the transfer of process-IP flow mappings. Both are performed when a socket is bound via the network stack API. To this end, the socket API is extended with driver calls performing the specific changes. Whenever a new process registers or frees a socket, the operating system transparently adjusts the number of queues and their parameters. Delay timers and queue size must be set to fit specific scenarios.%

The system must be dynamically adjustable during runtime to facilitate changes in processes or IP flows. With the configuration process being linked to the socket API, all necessary tuning parameters can be passed at any point in time by the registering process.%

As there is no explicit information about the receiving process in a network packet, there needs to be a mapping between packet meta data and processes. To this end, a map between IP flows and processes is created and placed on the NIC. In this design, the destination port is used to map a packet to a process and its priority.

\section{Evaluation}
\label{sec:evaluation}
This section outlines a first set of experiments conducted under increasing network loads and preliminary results. The experiments have been conducted using a simulation of the presented NIC design in combination with an ESP32 IoT device running FreeRTOS.

\subsubsection*{Experiments under High Load}

In all experiments the IoT device runs four worker processes of different priority. The processes are controlled over the widely used industrial communication protocol MODBUS/TCP\footnote{MODBUS/TCP traces provided by \cite{frazao2018denial}.}. 
As each of the processes binds its own socket, four queues with different interrupt moderation configurations are set up in the NIC.

One baseline experiment was performed without any interrupt moderation.
To observe the system under high traffic, it is subjected to packet floods ranging from 0 to 15000 packets per second. 
All experiments were performed four times using different absolute delay timer values for the added packet floods.
The values range from $800\,\mu s$ to $3200\,\mu s$ resulting in the designations \emph{nomod} (for unmoderated traffic), \emph{d800}, \emph{d1600}, \emph{d2400}, and \emph{d3200}. Each experiment runs for a duration of 30 seconds. 
We observed the progression of interrupts generated in respect to packets received and the additional runtime of the processes incurred by the network traffic, and their consequential deadline misses.

\subsubsection*{Results}

\begin{figure}
\centering
  \includegraphics[width=\linewidth]{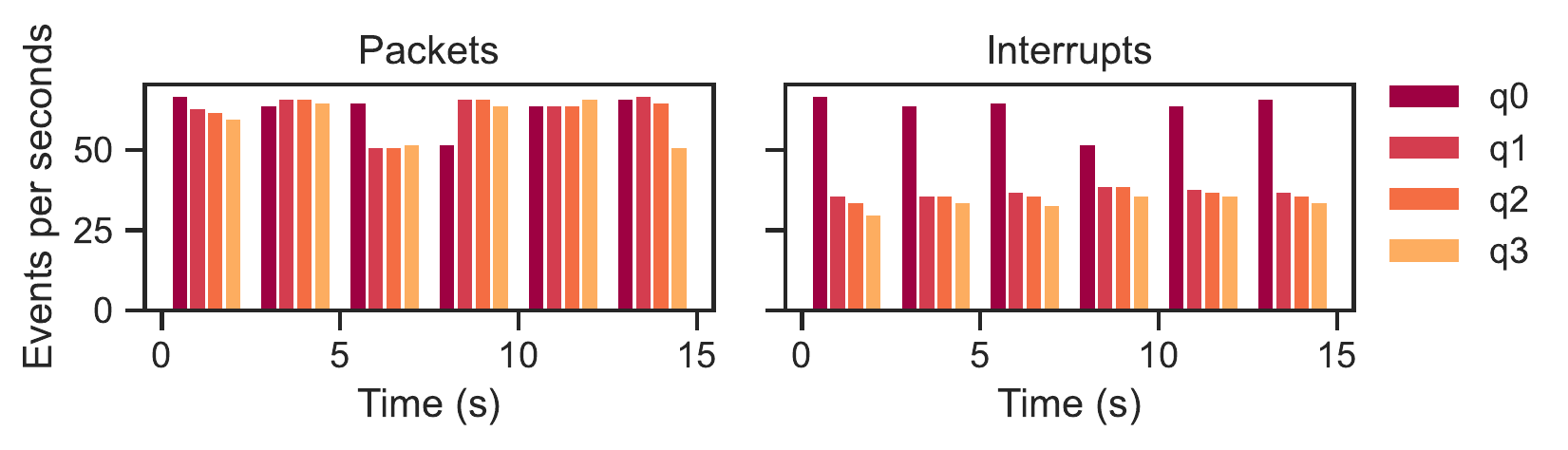}
  \caption{Packets and caused interrupts over time.}
  \label{fig:pckirq}
\end{figure}

Figure \ref{fig:pckirq} shows a comparison of packet and interrupt numbers for the baseline experiment without additional load. Queues 0 - 3 moderate interrupts in different time windows, so they generate fewer interrupts than the first queue, which is receiving packets for a critical task. The total rate of interrupts per packet ranged from 70\,\% in the undisturbed experiment to 2\,\% with high additional load of 15000 packets per second and $3200\,\mu s$ absolute timer value. The absolute moderation timer is an effective tool to moderate high loads as more packets are coalesced into interrupts while the critical task is unaffected.

\begin{figure}
\begin{minipage}{.46\linewidth}
  \centering
  \includegraphics[width=\textwidth,trim={0 0 0.1cm -0.1cm},clip]{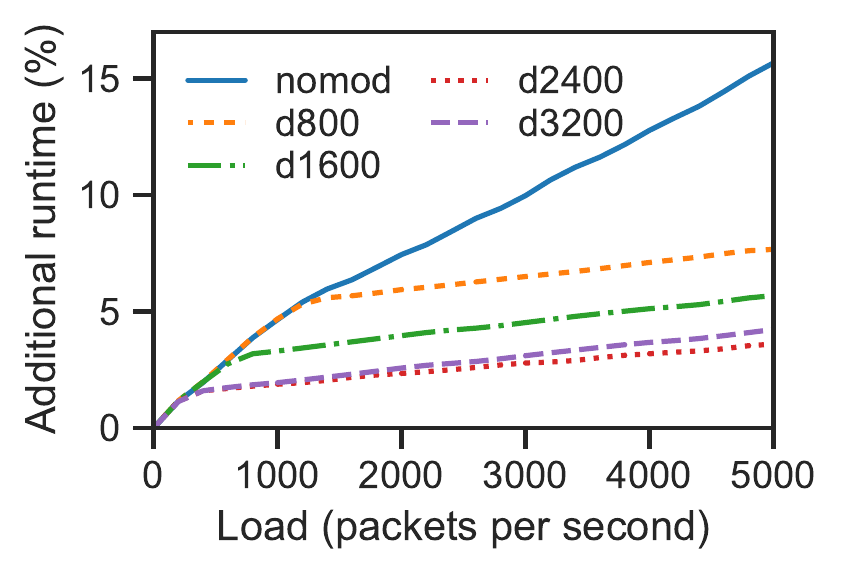}
  \caption{Additional runtime of critical process induced by increased load.}
  \label{fig:runtimeq0}
\end{minipage}%
\hspace{0.05\linewidth}
\begin{minipage}{.48\linewidth}
  \centering
  \includegraphics[width=\textwidth,trim={0.2cm -0.3cm 0 0cm},clip]{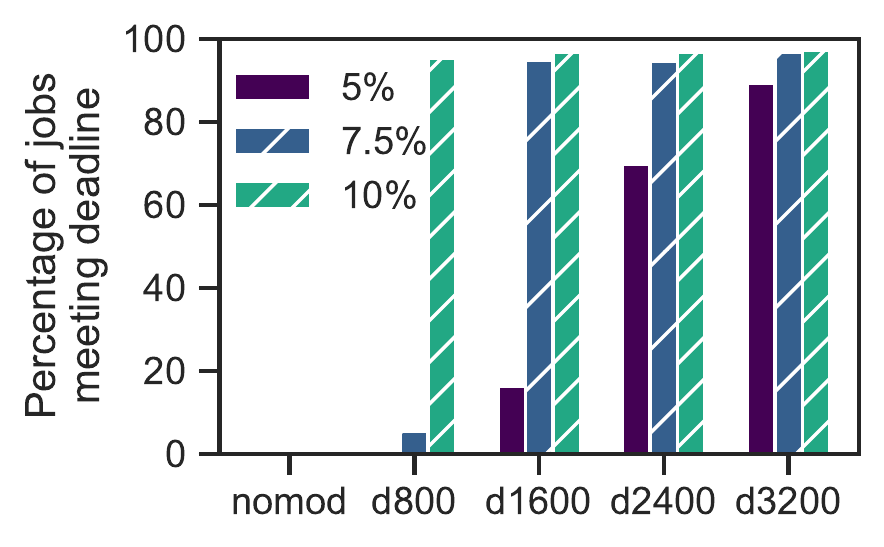}
  \caption{Share of critical packets meeting deadlines under load of 5000 pck/s.}%
  \label{fig:deadlines}
\end{minipage}
\end{figure}

Next, we observe the interrupt-induced runtime increase of the critical process. A significant mitigation of the malicious effects of packet floods can be obtained in all moderation configurations for the critical process. Processes of lower priority also benefit from the approach, as the CPU is freed up from unnecessary ISRs. Figure \ref{fig:runtimeq0} shows the mitigating effects for the critical task under variable additional load. Further, it can be seen that there is a scenario-specific optimal configuration between \emph{d2400} and \emph{d3200}. By increasing the delay parameters, more packets are coalesced for each interrupt, meaning that the networking tasks are confronted with larger bursts of packets per notification. This has negative effects on CPU load starting at a critical packet count. For the maximum depicted packet load of 5000 packets per second the additional runtime could be decreased by 80\,\% resulting from the prevention of 93\,\% of interrupts.

Figure \ref{fig:deadlines} shows the percentage of tasks for the critical process meeting their deadline with grace periods from 5\,-\,10\,\% compared to their baseline (median runtime without packet flood) under an additional load of 5000 per second.

\section{Related Work}
\label{sec:related}
The introduction of unpredictability in real-time environments through interrupts has been a long-standing research topic. In the following, we present past approaches to mitigate interrupt impact. %

The Advanced Interrupt Controller~\cite{gomes2015task} monitors the priority of the currently running process to determine if an interrupt should be triggered or held back by comparing it to the interrupt's priority. A simple extension of the interrupt controller unifies the priority spaces of attached interrupts and operating system processes. %

Network Interface Controllers with multiple transmit and receive queues have been introduced by Intel as early as 2007. The goal is to make use of multicore systems by parallelizing network load on the different queues. The trend is to increase the number of queues to facilitate cloud computing as Zhu et. al. showed in 2020~\cite{zhu2020data}. %

The priority inverting impact of interrupts in real-time systems has been identified and tackled by Amiri et. al. by employing priority inheritance protocols for interrupt service threads~\cite{amiri2015predictable}. %
In contrast, Multi-Sloth~\cite{multisloth} presents an OS adaptation that treats all threads as interrupts, scheduling threads and ISRs in a unified priority space.

The issue of DoS attacks in industrial IoT environments has been addressed by Niedermaier et al.~\cite{niedermaier2019secure}. A dual microcontroller architecture is proposed to separate networking tasks from critical real-time processes. %

\section{Conclusion}
\label{sec:conclusion}
Unexpected floods of network traffic can delay the process flow in real-time systems, which is a potential safety issue for many Industry 4.0 applications. To mitigate the effect of high traffic on real-time IoT devices, we propose a priority-aware multiqueue NIC that maps IP flows to processes when a socket is bound. %
We evaluated our design using a NIC simulation and an IoT device running a real-time operating system. The results of these experiments show that our approach significantly reduces the impact of traffic floods on critical process runtimes, saving 93\,\% of interrupts and 80\,\% of processing delay under packet rates of 5000 per second.%

\bibliographystyle{ACM-Reference-Format}
\bibliography{inc/bibliography} 

\end{document}